\documentclass[sigconf,screen]{acmart}

\newcommand{\asplossubmissionnumber}{127}
\usepackage[normalem]{ulem}
\usepackage{relsize}
\usepackage{adjustbox} % Auto-resize table content 
\usepackage{amsfonts}
\usepackage{amsmath}
\usepackage{comment}
\usepackage{enumitem}     % Allow customizations of itemize/enumerate env's
\usepackage{listings}
\usepackage{textcomp}     % Provides \textmu for upright mu's
\usepackage{xspace}
\usepackage{courier}
\usepackage{caption}
\DeclareCaptionFont{white}{\color{white}}
\DeclareCaptionFormat{listing}{\colorbox[cmyk]{0.43, 0.35, 0.35,0.01}{\parbox{0.48\textwidth}{\hspace{15pt}#1#2#3}}}
\usepackage{graphicx}
\PassOptionsToPackage{protrusion=true,expansion=true,kerning,spacing}{microtype}
\usepackage{units}        % For nice fractions, \nicefrac{1}{2} --> 1/2

\usepackage[version=4]{mhchem}

\usepackage[bb=boondox]{mathalpha}

\usepackage[capitalise,nameinlink,noabbrev]{cleveref}

\usepackage{array}
\newcolumntype{L}[1]{>{\raggedright\let\newline\\\arraybackslash\hspace{0pt}}m{#1}}
\newcolumntype{C}[1]{>{\centering\let\newline\\\arraybackslash\hspace{0pt}}m{#1}}
\newcolumntype{R}[1]{>{\raggedleft\let\newline\\\arraybackslash\hspace{0pt}}m{#1}}

\usepackage{amsmath,amsfonts}
\usepackage{algorithmic}
\usepackage{graphicx}
\usepackage{textcomp}
\usepackage{xcolor}
\PassOptionsToPackage{hyphens}{url}
\usepackage{subcaption}
\usepackage{multirow}

\usepackage{todonotes}
\def\BibTeX{{\rm B\kern-.05em{\sc i\kern-.025em b}\kern-.08em
    T\kern-.1667em\lower.7ex\hbox{E}\kern-.125emX}}

\title{Junkyard Computing: Repurposing Discarded Smartphones to Minimize Carbon}
\author{Jennifer Switzer, Gabriel Marcano, Ryan Kastner, and Pat Pannuto}
\usepackage[framemethod=TikZ]{mdframed}

\setcopyright{none}
\renewcommand\footnotetextcopyrightpermission[1]{}

\begin{document}
\begin{abstract}
1.5 billion smartphones are sold annually, and most are decommissioned less than two years later. Most of these unwanted smartphones are neither discarded nor recycled but languish in junk drawers and storage units. This computational stockpile represents a substantial wasted potential: modern smartphones have increasingly high-performance and energy-efficient processors, extensive networking capabilities, and a reliable built-in power supply. 
This project studies the ability to reuse smartphones as ``junkyard computers.'' 
Junkyard computers grow global computing capacity by extending device lifetimes, which supplants the manufacture of new devices. We show that the capabilities of even decade-old smartphones are within those demanded by modern cloud microservices and discuss how to combine phones to perform increasingly complex tasks.
We describe how current operation-focused metrics do not capture the actual carbon costs of compute.
We propose Computational Carbon Intensity---a performance metric that balances the continued service of older devices with the superlinear runtime improvements of newer machines.
We use this metric to redefine device service lifetime in terms of carbon efficiency.
We develop a cloudlet of reused Pixel 3A phones.
We analyze the carbon benefits of deploying large, end-to-end microservice-based applications on these smartphones.
Finally, we describe system architectures and associated challenges to scale to cloudlets with hundreds and thousands of smartphones.
\end{abstract}

\maketitle

\thispagestyle{plain}
\pagestyle{plain}

% Latex is dumb, move the tables here as commands so we can drop them in the paper for arrangement more easily

\newcommand{\tableMetrics}{
\begin{table*}
\centering
\caption{\textbf{The GeekBench performance on four benchmarks across the five devices.} \textmd{`Single' represents the single-core performance, and `multi' represents the multi-core performance. We consider the latter to be representation of the total computational power of the device. N is the number of devices that would be required to build a system with approximately the same computational power as a single PowerEdge R740 (baseline). This is calculated by dividing the PowerEdge multicore throughput by that of the reused device.}}
\begin{adjustbox}{width=1.00\linewidth}
\begin{tabular}{|lr|rrr|rrr|rrr|rrr|}
\hline
    & & \multicolumn{3}{c|}{SGEMM (Gflops)} & \multicolumn{3}{c|}{PDF Render (Mpixels/sec)} & \multicolumn{3}{c|}{Dijkstra (MTE/sec)} & \multicolumn{3}{c|}{Memory Copy (GB/sec)} \\
     & & Single & Multi & N & Single & Multi & N & Single & Multi & N & Single & Multi & N \\
     \hline
    PowerEdge R740 Server & 2017 & 77.2 & 2,070 & \textbf{1} & 109.1 & 3,140 & \textbf{1} & 3.58 & 80.2 & \textbf{1} & 6.33 & 19.5 & \textbf{1}\\
    HP ProLiant DL380 G6 & 2007 & 14.2 & 104.2 & \textbf{20} & 74.2 & 528.4 & \textbf{6} & 2.43 & 16.9 & \textbf{5} & 6.52 & 11.3 & \textbf{2} \\
    Lenovo Thinkpad X1 Carbon G3 & 2015 & 72.1 & 123.7 & \textbf{17} & 123.2 & 225.1 & \textbf{14} & 3.08 & 7.45 & \textbf{11} & 11.0 & 13.1 & \textbf{2} \\
    Pixel 3A Smartphone & 2019 & 8.84  & 39.0 & \textbf{54} & 38.9 & 147.0 & \textbf{22} & 1.08 & 4.44 & \textbf{19} & 4.00 & 5.45 & \textbf{6} \\
    Nexus 4 Smartphone & 2012 & 1.95 & 8.12 & \textbf{256} & 14.1 & 40.8 & \textbf{77} & 0.654 & 2.21 & \textbf{37} & 2.35 & 3.22 & \textbf{7}\\
   % \hline
    %\multicolumn{2}{|r|}{Resulting CCI Units} & \multicolumn{3}{c|}{CO2e/flop} & \multicolumn{3}{c|}{CO2e/Mpixel} & \multicolumn{3}{c|}{CO2e/MTE} & \multicolumn{3}{c|}{CO2e/GB} \\
    \hline
\end{tabular}
\end{adjustbox}
\label{tab:metrics}
\end{table*}
}

\newcommand{\tablePower}{
\begin{table}[h]
    \centering
        \caption{\textbf{Power (Watts) given CPU usage.}
    \textmd{Subscripts indicate CPU usage percentage. $P_{avg}$ calculated for light-medium workload.}}  %Measurements were performed by us, excepted for the PowerEdge, which is sourced from Dell's LCA \cite{dell_lca}.}
    \begin{adjustbox}{width=0.75\columnwidth}
    %jen redo pixel #s better later
    \begin{tabular}{|l|rrrrr|}
    \hline
        & $P_{100}$ & $P_{50}$ & $P_{10}$ & $P_{idle}$ & $P_{avg}$\\
        \hline
        PowerEdge & 510 & 369 & 261 & 201 & 308.7 \\ 
        ProLiant & 280 & 213 & 181 & 169 & 199.1 \\
        Thinkpad & 24 & 16.2 & 8.5 & 3.4 & 11.47 \\
        Pixel 3A & 2.5 & 1.9 & 1.4 & 0.8 &  1.54 \\
        Nexus 4 & 3.6 & 2.7 & 1 & 0.7 & 1.78 \\
        \hline
    \end{tabular}
    \end{adjustbox}
    \label{tab:power}
\end{table}
}

\section{Introduction}
\label{sec:introduction}
\label{sec:intro}

\begin{figure*}
    \centering
    \includegraphics[width=\textwidth]{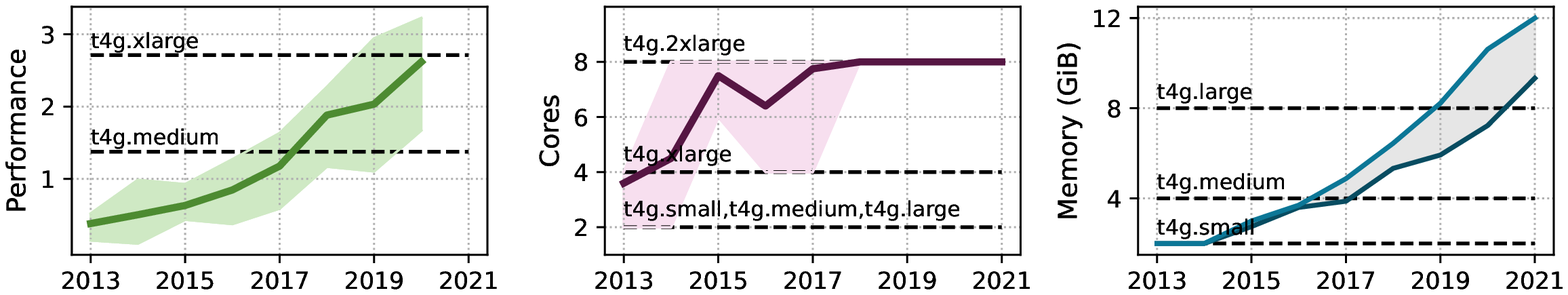}
    \caption{\textbf{The capabilities of recent smartphones meet or exceed that provided for modern microservices.}
    \textmd{The three plots show the performance (according to the GeekBench score\cite{bench_android}), number of cores, and memory for the five most popular Android phones released each year since 2013. A GeekBench score of 1 is equivalent to an Intel Core i3 processor. Solid lines indicate the mean. The shading shows the minimum and maximum ranges. The memory plot has two lines corresponding to the minimum- and maximum-memory configurations available to the consumer.
    The horizontal dotted lines show the capabilities of different Amazon EC2 T4g instance sizes as of August 2021, plotted for context. 
    }}
    \label{fig:motivation}
\end{figure*}

Manufacturing electronic devices is an energy-intensive process.
For devices with lower utilization, such as consumer-class electronics or over-provisioned servers, manufacturing dominates the lifetime carbon footprint~\cite{lca}.
This is especially true in devices with short use cycles.
In the United States alone, 150 million smartphones are discarded each year, amounting to one phone discarded per person every two years~\cite{150mil}.
As a result, manufacturing accounts for $70-85\%$ of the lifetime carbon footprint of a smartphone~\cite{lca,joshi2021life}.

Consumer electronics are also becoming increasingly powerful---the performance of recent smartphones rivals or exceeds that of an Intel Core-i3 processor (\Cref{fig:motivation}).
%~\cite{bench_android}.
%
Yet, phones are often discarded despite being completely (or partially) operational. Otherwise functional electronic devices are retired prematurely due to technical, style, or planned obsolescence~\cite{keeble2013obsolescence}.
Compare this to other high-priced devices with entire ecosystems of functional obsolescence, e.g., cars, which are resold until they are ``driven into the ground.'' 

Of course, the performance of computing devices improves much faster than that of a modern automobile.
What should the lifetime target be for compute devices?
Should we also run every electronic device ``into the ground?''
If we try, which applications should we run on these older, less-capable devices?
Can we simply aggregate devices to perform equivalent larger-scale computing?
And perhaps most importantly, how do we assess whether it is worth it: when are there carbon savings to be had by scavenging devices from the ``junkyard?'' 

To answer these questions, we introduce a new carbon-aware performance metric. \textit{Computational Carbon Intensity (CCI)} measures the lifetime carbon impact of a device versus the lifetime useful compute it performs.
CCI quantifies the value of extending the service lifetime of computational devices.

We apply CCI to old servers, old laptops, and old smartphones.
While each device type shows potential as carbon-saving hardware, we find that used smartphones (repurposed as general-purpose compute nodes) offer the best potential for carbon impact across a broader suite of potential applications.

Smartphones are an attractive target for repurposing for several reasons.
First, there is a remarkably large number of them.
Between 60-70\% of smartphones are neither thrown out nor recycled~\cite{baeckmanmobile,li2012survey}.
%
%Instead, they sit in drawers, which creates a massive stockpile of computational potential.
%
If even 10\% of the devices decommissioned in the last five years were available for repurposing, we would have 75 million new compute nodes.
Second, a smartphone comes with a wide array of valuable components: increasingly powerful processors, a robust uninterruptible power supply (batteries), and a diverse array of networking hardware, which includes local area connectivity (WiFi and Bluetooth) and long-range, high-performance uplink (cellular modems).
And they are tolerant of a wide range of environmental conditions, e.g., water-resistant or waterproof.

Can a smartphone meet the performance provided by purpose-built cloud servers? We argue that it can. \cref{fig:motivation} compares the performance, the number of cores, and memory in legacy smartphones to AWS T4g instances, which use Arm-based AWS Graviton2 processors. T4g instances target burstable general-purpose workloads like microservices, e-commerce platforms, small databases, development environments, and virtual desktops~\cite{aws-instance}. The analysis indicates that even a several-year-old smartphone could be valuable for this class of modern cloud services.

We put this idea to the test.
We benchmark discarded smartphones, some of which are now a decade old. We then build out a cloudlet of Pixel 3A smartphones.
These smartphones were purchased on eBay for \$65\,USD.
We measure the performance of the cloudlet using DeathStarBench, a benchmarking suite for cloud microservices~\cite{deathstarbench}.

The main contributions are:
\begin{enumerate}
    \item Defining CCI, a new metric to quantify the lifetime carbon impact of a device and our ability to amortize that impact (\cref{sec:metrics})
    \item Comparing the carbon and performance trade-offs of reusing old servers, laptops, and smartphones to identify the phone as a first-priority platform to explore for reuse (\cref{sec:metrics})
    \item Performing a design space exploration of how to architect legacy phones into a `junkyard computer' (\cref{sec:design})
    \item Building a real-world prototype cloudlet constructed from reused devices and benchmarking its performance using modern microservice applications from the DeathStarBench suite (\cref{sec:junkyard_cloudlet}).
\end{enumerate}
\section{Related Work}
\label{sec:related}
We are not the first to highlight the computational capability of wimpy cores~\cite{andersen2009fawn}, nor the first to suggest that phone-class hardware could be used~\cite{6217412,6290314}, or re-used~\cite{harizopoulos2011microcellstores,phones_compute}, in cloud computing contexts.
What differentiates this work is centering the design and evaluation on the amortized lifetime carbon impact and empirical testing on recovered hardware.

\subsection{Recycling is not the Answer}
\label{sec:background}
E-waste recycling of consumer electronics recovers less than 50\% of their materials ~\cite{VANEYGEN201653} and has a negative local impact on people and the environment~\cite{ewaste1,ewaste2}.
The inclusion of precious metals in ICs makes recycling them worthwhile, but the presence of toxic chemicals means that the recycling process is energy-intensive and hazardous~\cite{kumar}.
Wealthy countries in North America and the EU exacerbate global health disparities by shipping large amounts of E-waste to developing nations such as China and India. %%, where less regulation means higher profit margins. -- while this is very likely true, it strays from the point we are making here
Just three towns in China are estimated to process 11.5\% of the world's E-waste~\cite{guiyu1}.
Extensive studies of one of these towns revealed dangerous levels of pollutants in the surrounding environment~\cite{guiyu2} and elevated levels of heavy metal poisoning among residents.

\subsection{Computational Efficiency Metrics}
While efficiency metrics are useful for many applications, they do not capture the environmental and human cost of compute.
Greater energy efficiency does not always imply a lower carbon footprint, especially when efficiency is achieved via fast device turnover.

%%       Total Facility Energy       Non IT Facility Energy
%% PUE = --------------------- = 1 + ----------------------
%%        IT Equipment Energy          IT Equipment Energy

\textit{Power Usage Effectiveness (PUE)} is a standard metric for datacenter efficiency that expresses the ratio of the total energy consumption of the datacenter to the energy consumption of ICT devices alone, with 1.0 being an ideal value.
While PUE captures the operational efficiency of the datacenter's computational devices and cooling system, it does not reflect the tremendous energy that went into manufacturing the facility or the equipment housed within.

To understand why this is important, consider two datacenters, A and B.
For the same computational output, datacenter B consumes 10\% more energy than A, i.e., $\textnormal{PUE}_{(B)}=1.1\times\textnormal{PUE}_{(A)}$.
However, datacenter A achieves this efficiency by upgrading its servers at a rate that is $2\times$ faster than datacenter B; i.e., datacenter A's ICT manufacturing costs are $2\times$ that of datacenter B.
Let us assume that manufacturing is responsible for 20\% of the carbon emissions associated with both datacenters and that operational energy accounts for the other 80\%.
Then, the carbon footprint of datacenter B compared to that of datacenter A is:
\begin{align*}
 % Same math, just factored so that eqn matches text
 % \ce{CO_2}e_{(B)} &= (0.8*1.1+0.2*0.5)\times\ce{CO_2}e_{(A)} \\
  \ce{CO_2}e_{(B)} &= 0.8*(1.1\times\ce{CO_2}e_{(A)})+0.2*(0.5\times\ce{CO_2}e_{(A)})
  \\
  \ce{CO_2}e_{(B)} &= 0.98\times\ce{CO_2}e_{(A)}
\end{align*}
Where $\ce{CO_2}e$ is the Global Warming Potential (GWP), or carbon footprint, in units of $\ce{CO_2}$-equivalent weight.
Despite having a higher PUE, datacenter B has a slightly lower carbon footprint because it uses its servers longer.

\textit{Total Cost of Ownership (TCO)} amortizes initial purchase costs and captures runtime efficiency via operational overhead. However, as we will show, these economic costs are not always aligned with carbon costs.

\subsection{Characterizing the Carbon Footprint of Computation}

Patterson et al. argue that researchers should consider the carbon emissions incurred by machine learning training~\cite{patterson}. Their formula for the carbon emissions of training is similar to our definition of compute carbon ($\mathbb{C}_{C}$, introduced in the next section) with the exception of scope: we calculate operational carbon for the entire lifetime of a system as opposed to a single model training event. In both cases, the energy source is important to the overall carbon footprint. Patterson et al.\ do not address embodied carbon.

Gupta et al.~discuss the importance of accounting for embodied carbon~\cite{chase}. Like us, they note that a significant fraction of smartphone carbon emissions come from manufacturing. They further note that recent optimizations have focused on maximizing performance rather than considering carbon footprint. Like us, they argue for longer device lifetimes: they state that amortizing the carbon footprint of mobile devices requires continuously operating them for at least three years. In follow-on work, Gupta et al. develop carbon-conscious metrics to enable architectural-level modeling of carbon consumption and describe the carbon benefits of repurposing smartphones as an SSD storage system~\cite{gupta2022act}.

\emph{Life-cycle assessment (LCA)} seeks to characterize the lifetime environmental impact of a product across multiple metrics, including carbon. LCAs have been performed for smartphones, laptops, and servers \cite{lca,poweredge}. Manufacturing accounts for $70-90\%$ of the lifetime carbon footprint for phones and laptops.
For high-performance computing devices, manufacturing accounts for $20-50\%$ of the lifetime carbon cost~\cite{poweredge,lca,joshi2021life}. %These numbers are expected to increase as datacenter operators transition to renewable energy sources, which will decrease the carbon intensity of operations, thus increasing the fractional cost of manufacturing~\cite{chase}. 
LCAs measure the total carbon footprint of a device across its lifetime; our approach (CCI) is different in that we consider the amortized carbon footprint of each unit of computational work. We use vendor-provided LCA numbers to parameterize our CCI calculation.

Raghavan and Ma characterize the \emph{embodied energy} of the internet~\cite{emergy}.
Subsequent work argues for the importance of including embodied energy in discussions of sustainable computing more generally~\cite{pargman2014rethinking}.
With this work, we attempt to take a step towards fulfilling this mandate by providing architects and systems engineers the tools to express embodied energy quantitatively.

\subsection{Repurposing Consumer Electronics for HPC}

Rajovic et al.\ propose using mobile SoCs for high-performance computing (HPC)~\cite{hpc}.
Their analysis finds that mobile SoCs are sufficiently performant for many applications and are more energy efficient than traditional HPC chips.
Their follow-on work builds a mobile SoC-based cluster~\cite{tibidabo}.
Their implementation uses new and isolated mobile SoCs and is not focused on reuse.

Shahrad and Wentzlaff propose a server built from decommissioned mobile phones~\cite{phones_compute}.
While the work considers E-waste reduction as motivation, it does not look into quantifying the carbon impact.
They present a design proposal but do not include an implementation or empirical evaluation.

B\"usching et al.\ empirically test six Android phones over WiFi to evaluate cluster performance on LINPACK~\cite{droid}.
They do not consider other workloads, architectures, or management costs.

% Just here for layout
\tableMetrics
%\tablePower

\subsection{Repurposing Smartphones as Sensors}
% %

An orthogonal line of work is using smartphones in IoT and sensor networks ~\cite{pp_android,zink2014comparative}.
Mobile phone operating systems are optimized for interactive use, and when human interaction is completely removed, a very long tail of deployment-ending issues crops up.
This experience led us to replace the standard mobile phone operating system with a more traditional technology stack for our smartphone-based system.

\section{Quantifying Carbon}\label{sec:method}%idek--jen
\label{sec:metrics}

The first step in repurposing devices is assessing their performance and carbon consumption tradeoffs versus a new system. Can we quantify the carbon benefit of repurposing devices? How much performance are we giving up by not using the latest technology?  Which device provides the best tradeoff between performance and environmental impacts?

We study the performance and carbon benefits of building a server using different legacy devices. Our baseline is a PowerEdge R740---a modern server for which Dell has published a full LCA~\cite{dell_lca}. We aim to replicate the R740 functionality with repurposed devices. We compare four building blocks: a 15-year-old server (HP ProLiant DL380 G6), an 8-year-old laptop (Lenovo Thinkpad X1 Carbon G3), a decade-old smartphone (Nexus~4), and a 3-year-old smartphone (Pixel 3A).

\subsection{Workload}
We assume that the devices operate at the light-medium operating regime specified in Dell's PowerEdge R740 LCA \cite{dell_lca}. This operating regime has the following load profile:
\begin{itemize}
\item 100\% load mode: 10\% of the time.
\item 50\% load mode: 35\% of the time.
\item 10\% load mode: 30\% of the time.
\item Idle mode: 25\% of the time.
\end{itemize}
We further consider what supporting datacenter infrastructure must be added to each configuration to provide the following:
\begin{itemize}
    \item 1\,Gbps networking capacity.
    \item 30\,minutes backup power.
    \item Sufficient cooling to operate in 25\,C ambient temperature.
\end{itemize}
\subsection{Performance} 
\Cref{tab:metrics} shows the performance of the five different devices across four applications from the Geekbench suite.\footnote{We generally eschew performance benchmarks such as SPEC due to their unavailability and inaccessibility for mobile platforms. To date, Geekbench is the most commonly used measure in studies that focus on a broader array of platforms, particularly those which include mobile platforms~\cite{chase, phones_compute}.}
The PowerEdge Server has the highest performance, followed by the old server, old laptop, and old smartphones. $N$ provides the performance ratio between the PowerEdge Server and other devices. That value changes depending on the application. For example, there is a $256\times$ difference between the PowerEdge Server and the Nexus 4 for SGEMM, while the relative performance difference is just $7\times$ for Memory Copy.

\subsection{Energy Consumption}
Device power draw plays a prominent role in the overall carbon footprint. We perform a CPU stress test on all devices to characterize their power draw in different regimes. We turn off WiFi and Bluetooth on the phone and laptop for the stress test. We turn off the phone screens and turn the laptop's brightness as low as possible. We run the Linux \verb+yes+ command \verb+n+ times, where \verb+n+ is the number of cores the device has and use \verb+cpulimit+ to restrict the CPU usage of each \verb+yes+ process. The phone’s power draw is measured via a USB power meter, which sits between the device and the power source. %\footnote{Note that while the device can draw power from the battery while plugged in, we did not observe this happening. The device was fully charged before and after the test.}. 
The laptop measurements use a plug-through power meter. We query the HPE iLO interface for the ProLiant servers.

\Cref{tab:power} presents a summary of the results. Not surprisingly, the servers (PowerEdge, ProLiant) have the highest power draw, followed by the laptop and smartphones. 

\tablePower

\subsection{Carbon Cost}
To quantify the environmental cost of proposed solutions beyond operational energy, we need new metrics that properly capture the carbon footprint over the system's lifetime.
An \emph{ideal carbon metric}:
%\smallskip
\begin{enumerate}
    \item Rewards operational energy efficiency.
    \item Rewards manufacturing efficiency.
    \item Rewards the reuse of already-manufactured devices.
    \item Reflects computational work achieved per unit carbon.
\end{enumerate}
\smallskip
Existing efficiency metrics, e.g., PUE, fulfill point \#1 but fail to capture points \#2-4. Lifetime carbon footprint---a measure of the greenhouse gas emissions of a device over a standard lifetime, reported by companies like Apple~\cite{ipad,macbook} and Dell~\cite{dell_lca, poweredge}---does not satisfy point \#4. Thus, we must define a metric that captures all four points.

\begin{figure*}
    \centering
    \includegraphics[width=0.33\textwidth,trim=0.1cm 0.1cm 0.1cm 0.1cm]{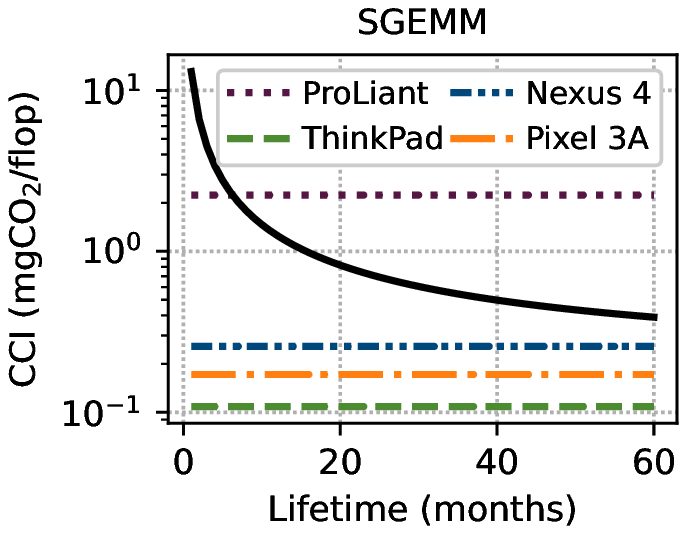}%
    \includegraphics[width=0.33\textwidth,trim=0.1cm 0.1cm 0.1cm 0.1cm]{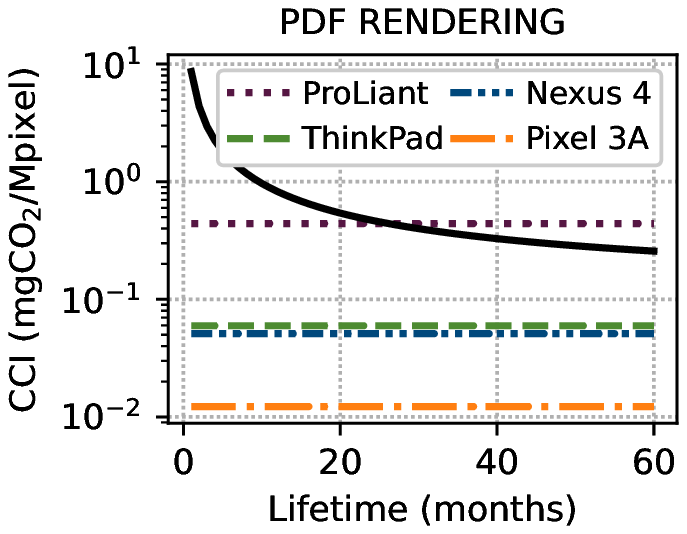}%
    \includegraphics[width=0.33\textwidth,trim=0.1cm 0.1cm 0.1cm 0.1cm]{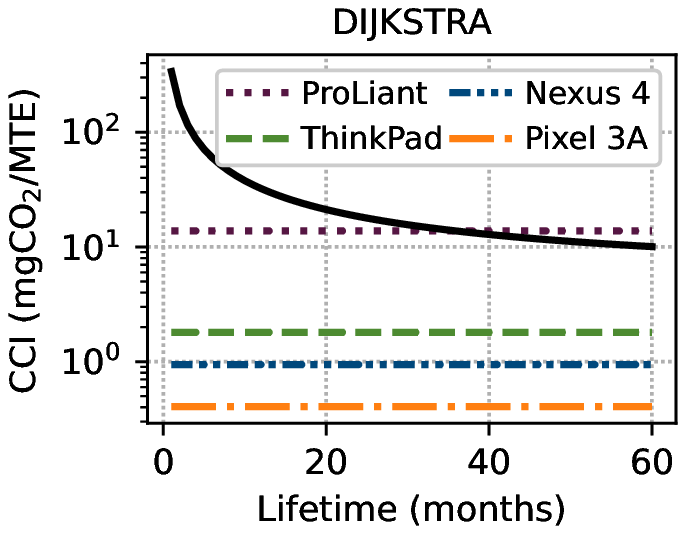}
    \caption{\textbf{Single-device CCI trends for three benchmarks.}
    \textmd{Lower CCI values are better. This analysis assumes no peripherals are needed and a California energy mix.
    The chosen benchmark affects the relative performance of each device.
    In all cases, the CCI of the reused smartphones and laptop is lower than that of a new server.
    The reused server does not perform as well due to its poor \nicefrac{performance}{Watt}.}}
    \label{fig:cci_simple}
\end{figure*}

\emph{Computational Carbon Intensity (CCI)} is the \ce{CO_2}-equivalent released per unit of computation work.
We calculate this metric across the entire lifespan of our devices to get their amortized carbon intensity.
The general formula is as follows:
\begin{equation}
    \textnormal{CCI} = \frac{\sum\limits_{lifetime}\ce{CO_2}e}{\sum\limits_{lifetime}\textnormal{ops}}
\end{equation}
The numerator can be further broken up into the carbon associated with manufacturing, compute, and networking:
\begin{equation}
    \textnormal{CCI} = \frac{\mathbb{C}_{M}+\mathbb{C}_{C}+\mathbb{C}_{N}}{\sum\limits_{lifetime}\textnormal{ops}}
\end{equation}

$\mathbbb{C}_{M}$ is the carbon associated with the manufacture of the device (often called the ``embodied carbon'').
%
%We calculate this value following the methodologies presented in \cite{lca} and include the carbon cost of all devices and peripherals.
%
It is a single upfront cost at the beginning of the cluster's lifetime. We source our values for $\mathbbb{C}_{M}$ from the life cycle assessment literature.

%\smallskip
%\noindent
$\mathbbb{C}_{C}$ is the carbon associated with compute.
Equivalently, it is the total carbon footprint of the energy required to perform the computations over the lifetime of the device, calculated as:
\begin{equation}
    \mathbb{C}_{C} = \sum\limits_{lifetime}\textnormal{CI}_{\textnormal{grid}} * E
\end{equation}
where $\textnormal{CI}_\textnormal{grid}$ is the carbon intensity of the grid, in units of $\frac{\ce{CO_2}e}{\textnormal{joule}}$, and $E$ is the cluster's energy consumption, in joules. At the light-medium workload, this works out to:
\begin{equation}\label{eq:cc}
\begin{split}
    \mathbb{C}_{C} = \textnormal{CI}_\textnormal{grid} * \sum\limits_{lifetime} 0.10*P_{100\%}\ +\ 0.35*P_{50\%} \\
        +\ 0.30*P_{10\%}\ +\ 0.25*P_{idle}
\end{split}
\end{equation}

%\smallskip
%\noindent
$\mathbbb{C}_{N}$ is the carbon associated with networking:
\begin{equation}\label{eq:net}
    \mathbb{C}_{N} = \sum\limits_{lifetime} \textnormal{CI}_\textnormal{grid} * f_\textnormal{net} * EI_\textnormal{net}
\end{equation}
where $f_\textnormal{net}$ is the rate at which data is sent and received, in $\frac{\textnormal{bytes}}{\textnormal{second}}$, and $EI_\textnormal{net}$ is the energy intensity of networking, in $\frac{\textnormal{joule}}{\textnormal{byte}}$.
CCI sums these three carbon components and divides by the total number of operations computed in the device's lifetime. %

CCI satisfies point \#1 as it accounts for operational energy consumption per instruction in the numerator ($\mathbb{C}_{C}$).
It satisfies point \#2 via the inclusion of $\mathbb{C}_{M}$.
The total carbon cost is amortized per instruction; thus, point \#4 is satisfied.
To satisfy point \#3, we stipulate that when reusing a device, the carbon cost of manufacturing is considered already paid, i.e., $\mathbb{C}_{M}=0$.

\paragraph{CCI in practice}With this definition, we can do an initial, per-device CCI calculation for the servers, laptop, and phones.
 
$\mathbb{C}_{C}$ is calculated per \Cref{eq:cc} using the $P_{avg}$ from \Cref{tab:power} and the throughput from \Cref{tab:metrics}. We do not include the networking term $\mathbb{C}_{N}$ for this single device analysis. As the reused devices are already manufactured, we set their $\mathbb{C}_{M}$=0. 

\emph{Operations per second (ops)} varies depending on the computational work done. For example, the GeekBench SGEMM benchmark is measured in floating point operations, PDF Rendering uses pixels, Dijkstra computes pairs, Memory Copy is GB, etc.~\cite{geekbench4}. CCI depends on the type of operation; thus, its units can change depending on the benchmark under consideration. Operations are calculated as follows: We scale the throughput achieved in the micro-benchmark under consideration by the CPU usage to get operations per second, then multiply by the lifetime to get the total operations computed. 

We assume that throughput scales linearly with CPU usage, e.g. $\textnormal{ops}_{50\%}\ =\ 0.5*\textnormal{ ops}_{100\%}$. This simplification is necessary when extrapolating from microbenchmarks; when benchmarking a full system (\Cref{sec:junkyard_cloudlet}), we no longer make this assumption. Under the light-medium workload described previously, the average operations per second (ops) is calculated as:
\begin{equation}
\begin{split}
    \textnormal{ops}_{avg} = 0.10*\textnormal{ops}_{100\%}\ + 0.35*\textnormal{ops}_{50\%}+ 0.30*\textnormal{ops}_{10\%}
\end{split}
\end{equation}

\Cref{fig:cci_simple} shows the CCI over the lifetime of the five devices. Lower CCI values are better. Only the PowerEdge server incurs manufacturing carbon cost $\mathbb{C}_{M}$; repurposed devices have $\mathbb{C}_{M}=0$. Generally speaking, the phones have the best (lowest) CCI except in the SGEMM benchmark. Here, the specialized computational hardware available on the laptop outweighs its higher power draw.

\paragraph{Assumptions and limitations}
It is difficult to account for every possible impact of a computing system. For instance, CCI, as it stands, does not account for the carbon footprint of human labor and transportation beyond initial manufacture. 
%by the addition of a term to the numerator.

For our analysis, we assume that any device(s) being repurposed are essentially `free' in terms of manufacturing carbon, i.e.,  $\mathbb{C}_{M} =0$.\footnote{Although as we will see below, when building more complex systems the carbon cost of added peripherals must be considered.} An alternate analysis considers the initial manufacturing cost of the repurposed device, amortized by the work done in its first life. 
This alternate CCI formula is:
\begin{equation}
    \textnormal{CCI} = \frac{\mathbb{C}_{M}+\mathbb{C}_{C(1^{st}\  life)}+\mathbb{C}_{N(1^{st}\  life)}+\mathbb{C}_{C(2^{nd}\  life)}+\mathbb{C}_{N(2^{nd}\  life)}}{\sum\limits_{1st\ lifetime}\textnormal{ops(1}^{\textnormal{st}}\textnormal{  life)}+\sum\limits_{2nd\ lifetime}\textnormal{ops(2}^{\textnormal{nd}}\textnormal{  life)}}
\end{equation}
\noindent While this alternate form might also be helpful for certain analyses, we find that it is difficult to reason about the operational carbon footprint and operations computed in the device's `first life.' In our case, we are working with used devices from eBay and have no visibility into their first lives.

Another limitation of CCI is that it focuses on computing and does not capture other types of reuse, e.g., it does not account for whether or not the camera, microphone, and networking capabilities are being reused. Consumer devices contain many subcomponents supporting different compute requirements: CPUs, GPUs, custom hardware accelerators (audio, encryption, AI), networking hardware, power supplies, and batteries.
Ideally, every component is reused, but in reality, not all are required. For example, devices in a server role may not need a display or audio components. 

\emph{Reuse Factor} is a metric to account for the usage of the subcomponents. It weighs each component by its carbon footprint to estimate the embodied carbon of the components of the device that are being reused. It is calculated as:
\begin{equation}
  \textnormal{RF} = \frac{\sum\limits_{reused}\mathbb{C}_{M(i)}}{\mathbb{C}_{M}}
\end{equation}

\Cref{tab:working_estimates} gives our best attempt for the carbon intensity of the smartphone components. Coarse-grained ratios are sourced from \cite{lca}, which reported a 77\% contribution from ICs, 10\% contribution from the display, 3\% from the battery, and the rest from other components, including the PCB. The contribution from ICs was then further broken down using images from a Nexus 4 teardown and assuming that the relative contribution of each chip scales according to its size.

It is difficult to find accurate numbers for the fractional contributions of each smartphone subcomponent to the overall embodied carbon. Better embodied carbon estimates are necessary to create more accurate models. %\footnote{
The ACT is a tool for modeling architectural carbon footprint~\cite{gupta2022act}. % we aim to incorporate in future work. 
Currently, ACT only has estimates for ICs, which align with our estimates here, but does not consider all elements of the phone (e.g., battery, display, PCB, and chassis, which contribute  \textasciitilde{}\nicefrac{1}{3}).
%to be the only measure we consider for our analysis.
Thus, ACT is complementary to our work, and we are excited by the potential of its component models to improve our more holistic models.
%}

The reuse factor should be treated as a proxy for the extent to which a device has been reused. 
To see how the reuse factor is used, consider a cloudlet example: A smartphone is repurposed as a compute node. It is network-connected and is always in use. Code and results may use storage on the device and its battery as a UPS.
In this scenario, the compute capabilities, networking, battery, and storage are all reused, while the display and sensors are not. This yields a reuse factor of 0.85.

\begin{table}[]
    \centering
    \caption{\textbf{Working estimates for the fractional carbon intensity of various Nexus 4 subcomponents.}} %hdmi?
    \begin{adjustbox}{width=\columnwidth}
    \begin{tabular}{|l|r|r|p{3.9cm}|}
    \hline
    \textbf{Category} & \multicolumn{2}{|c|}{\textbf{Contribution}} & \textbf{Includes}\\
    \hline
        Compute & 25\% & 12.5\,kgCO$_{2}$ & \smaller 2GB RAM (Samsung K3PE0E00A), Snapdragon S4 Pro \\
        Network & 15\% & 7.5\,kgCO$_{2}$ & \smaller LG and WiFi chips: Qualcomm MDM9215M, Murata SS2908001, Qualcomm WTR1605L, Broadcom 20793S\\
        Battery & 15\% & 7.5\,kgCO$_{2}$ & \smaller Battery, power management (Qualcomm PM8921, Avago ACPM-7251) \\
        Display & 10\% & 5\,kgCO$_{2}$ & \smaller Screen \\
        Storage & 10\% & 4\,kgCO$_{2}$ & \smaller 8GB flash \\
        Sensors & 5\% & 3\,kgCO$_{2}$ & \smaller Accelerometer, camera, audio codec \\
        Other & 20\% & 10\,kgCO$_{2}$ & \smaller PCB, Chassis, packaging, other ICs\\
        \hline
    \end{tabular}
    \end{adjustbox}
    \label{tab:working_estimates}
\end{table}

\section{Designing the Junkyard}
\label{sec:design}
We showed that repurposed smartphones have the potential to outperform a traditional server in terms of carbon per op.
This section discusses how to transform discarded consumer smartphones into general-purpose compute nodes. We aim to answer the question: What does it take to make a server out of smartphones? 

\begin{figure}
\subfloat[\centering 100\% load.]{\includegraphics[width=.248\textwidth]{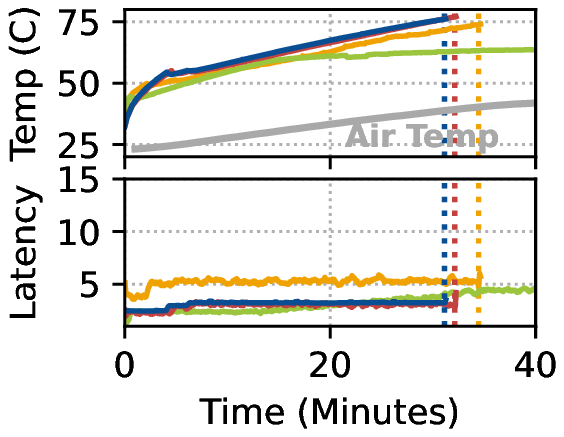}}
\subfloat[\centering Light-medium workload.]{\includegraphics[width=.248\textwidth]{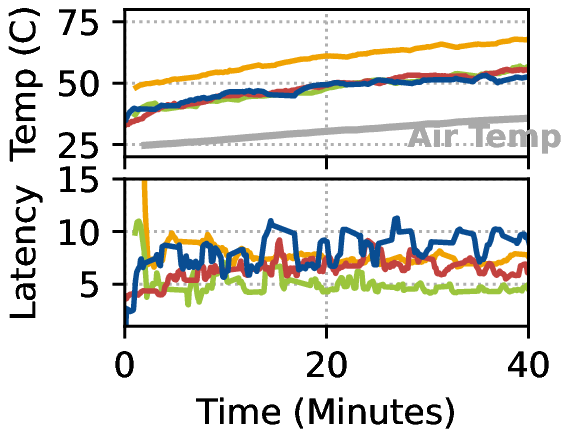}}
\caption{\textbf{Results of a thermal stress test.}
\textmd{The top graph shows the temperature internal to each phone (colored lines) and the external air temperature (grey line). The bottom graph shows the test job latency. For both graphs, the green line is the Nexus\,5; other colored lines are the Nexus\,4s. The horizontal dotted lines at the ends of the Nexus\,4 curves represent the device shutting itself off.}}
\label{fig:thermal}
\end{figure}

\subsection{Cooling}
Datacenter cooling can account for as much as 40\% of operational energy consumption~\cite{cooling}.
While we envision cloudlets rather than warehouses as a more realistic outcome for repurposed phones, we still anticipate packing heat-generating compute into limited space.

Phones are designed not to burn your hand when you use them, which means they have a strict upper limit on their thermal dissipation.
They begin throttling around 40-50\textdegree{}C \cite{kang2019fire}. Thus, a well-behaved device should never overheat due to its operation.
%However, it is possible to overheat due to high ambient temperature.
At around 60-70\textdegree{}C, the device will shut itself off to protect the user~\cite{kim2019thermal}. 

While collecting many smartphones in an enclosed space can create a high ambient temperature, we postulate that the individual thermal throttling of each device has the aggregate effect of limiting the air temperature reached. %
To test this hypothesis, we perform an experiment.

\paragraph{Experimental Set-Up} Four Nexus 4s and one Nexus 5 are placed in a sealed $5\times15\times10.5$\,inch Styrofoam box.
We run two scenarios: a CPU stress test, with a 90\%+ CPU job running on all four cores, and a simulated light-medium workload.
We log job throughput, device internal temperature, and air temperature in the box.

\paragraph{Results}
\Cref{fig:thermal} summarizes performance and temperature results.
We also calculate thermal power for both scenarios:\footnote{%
We assume the phones can be modeled as a block of silicon and that the temperature of the phones and the air was uniform throughout the medium.
}
\vspace{-1mm}
\begin{equation}
\begin{split}
    P_{T} = \frac{c_{p(air)} * m_{air} * \Delta T_{air}}{\Delta time} + \frac{\sum\limits_{phones} c_{p(Si)} * m_{ph.} *  \Delta T_{ph.}}{\Delta time}
\end{split}
\end{equation}
We calculate the thermal power prior to the shutdown of any devices. 
Total thermal power for the 100\% load scenario is \textasciitilde{}13\,W (2.6\,\nicefrac{W}{device}), and \textasciitilde{}6\,W (1.2\,\nicefrac{W}{device}) for the light-medium scenario.

Some additional observations:
a) The phones shut off at an internal temperature of 75-80\textdegree{}C. For most devices, this happened at an air temperature of 40\textdegree{}C.
b) The Nexus\,5 did not overheat in either scenario.
c) As the temperature increased, performance decreased.
d) Even under a 90\%+ (CPU) load, the phones exhibited a thermal power (2.6\,W) well below their 5\,W thermal design point (TDP).

At the cloudlet scale, with 256 Nexus 4s running at 100\%, we expect up to 666\,W of thermal power. This is within the cooling abilities of two COTS server fans rated for 500\,W, which would add 4\,W of power per fan~\cite{server_fan}. A rough estimate based on weight \cite{embodied_weight}, and assuming a world energy mix during production, yields an embodied carbon of 9.3\,kg per fan. We consider this added embodied carbon when estimating the carbon efficiency of cloudlet-scale smartphone clusters in \cref{sec:cloudlet}.

\paragraph{Scaling Further}

Aggressive thermal management reduces the cooling demand of a hypothetical smartphone datacenter but also limits its performance potential.
A phone's modest thermal dissipation capacity means that its effective sustained TDP is much lower than its transient peak TDP.
Thus, phones are well-suited to bursty workloads, but a poorer fit for more sustained compute without an aggressive adaptation of their thermal management systems.

\subsection{Networking}\label{sec:network}
There are many ways to network the phone cluster.
We consider two general use cases:
(1) the in-situ, edge compute case, where the cluster operates independently with access only to the cellular network
and (2) the existing-infrastructure case, where the cluster has access to a pre-existing WiFi or wired network.
The first case reflects applications in remote environments. The second might occur if the cluster was housed within an office building or datacenter.

Na et al.\ found that networking a co-located smartphone cluster over WiFi is not feasible; interference made wireless networking intractable beyond 30 devices~\cite{na2021scalable}. Thus, in a datacenter-scale scenario, we expect that wired clusters of smartphones connected to network switches would be employed. 

For smaller clusters operating away from external infrastructure, we propose a tree topology, with phones organized into sets of five, each with one hotspotted device. The hotspotted device communicates with the outside world via LTE and other devices via its WiFi network. WiFi is the more bandwidth limiting, e.g., the Nexus 4/5 smartphones have 150\,Mbit/s for both the uplink and downlink~\cite{old_wifi}. The tree topology leads to a capacity of 18.5Mbit/s downlink and uplink per device. While wired networking is best as the cluster scales, for smaller clusters (i.e., our ten-device cloudlet described in \Cref{sec:junkyard_cloudlet}), we have found a wireless network to be sufficient.

\subsection{Batteries \& Smart Charging} A critical advantage of smartphones is their built-in batteries. By charging during periods of higher renewable (``greener'') energy production, we can reduce the operational carbon intensity of the cluster. We call this \textit{smart charging.}

Smart charging opportunistically charges the devices whenever the grid-level carbon intensity falls below a tunable threshold. This threshold is based on historic grid characteristics and the charged device. We devised a heuristic for the California grid, but other regimes might require a different strategy. The device is charged if the battery level drops below 25\%, regardless of grid conditions. 

\begin{figure*}[htp]
\centering
\subfloat[\centering Carbon intensity in California.]{\includegraphics[width=.3\textwidth]{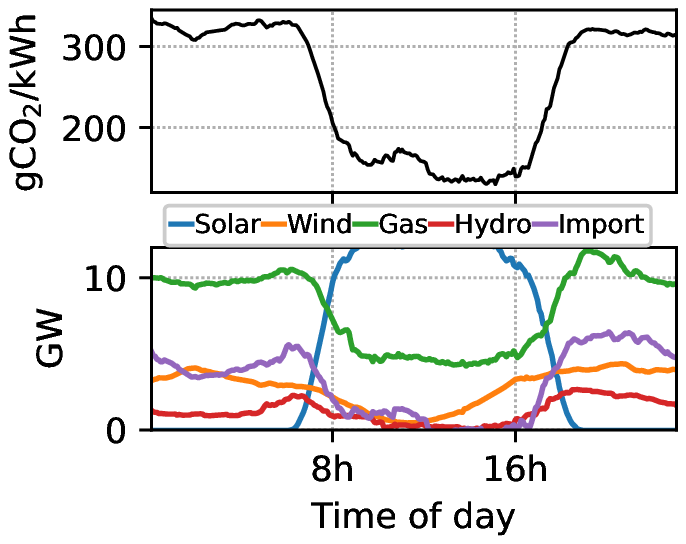}}
\subfloat[\centering Pixel 3A.]{\includegraphics[width=.3\textwidth]{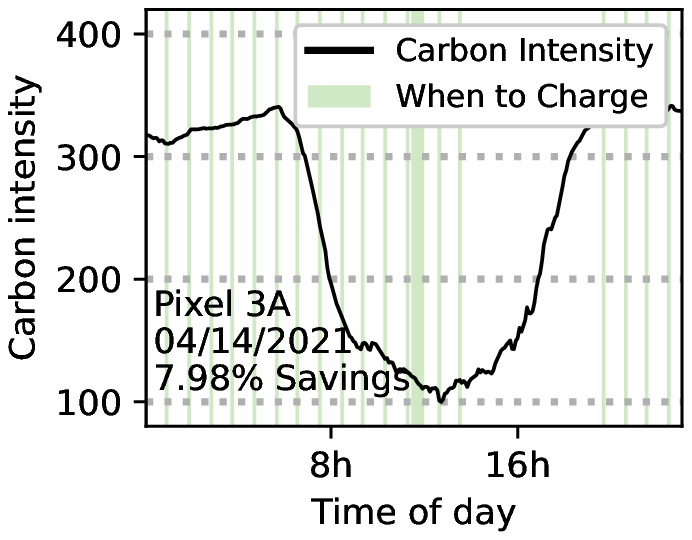}}
\subfloat[\centering ThinkPad X1 Carbon Gen 3.]{\includegraphics[width=.3\textwidth]{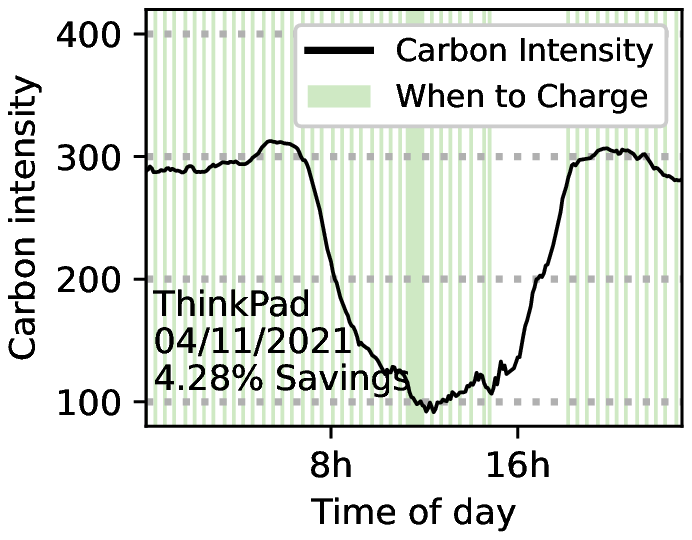}}
\caption{\textbf{Smart charging saves carbon.}
\textmd{Results are shown for a representative day during the month of April 2021. The black curve indicates the carbon intensity of the California grid throughout the day, and the green shading gives the charging periods suggested by the smart charging algorithm. On this particular day, the Pixel 3A's carbon emissions are reduced by 7.98\%, and the ThinkPad's carbon emissions are reduced by 4.28\%. Over the course of the month as a whole, the Pixel 3A sees median carbon savings of 7.22\%, with a standard deviation of 5.93\%. The ThinkPad sees median carbon savings of 4.03\%, with a standard deviation of 2.2\%}}
\label{fig:smart_charging}
\end{figure*}

We evaluate this algorithm on publicly available data from the California Independent System Operator \cite{caiso_data}. In California, the carbon intensity of the grid tends to be anti-correlated with solar production trends (\Cref{fig:smart_charging}a). Since this pattern is relatively uniform across days, we set the threshold as the Pth percentile of the previous day's instantaneous carbon intensities, where P is the percent of time the device will need to spend charging.

In California, the smart charging algorithm schedules charging when solar production is highest (middle of the day). The carbon savings vary depending on the device's battery capacity and charging rate. The Pixel 3A, with a 3\,Ah battery and 18\,W charging rate, sees a median carbon reduction of 7.22\% for the time period studied, while the ThinkPad X1 Carbon Gen3 laptop sees a median 4.03\% reduction during the same time period. The laptop sees smaller savings than the Nexus phone because its much higher power consumption (11.4\,W versus 1.54\,W) offsets its larger battery capacity.

Batteries also provide a convenient source of backup power. For a Pixel 3A operating on a light-medium workload, a 25\% charge should last slightly under 2 hours.
The 25\% minimum can be increased to provide a higher margin. Conversely, the minimum charge threshold can also be decreased to prioritize carbon savings over backup power.

There is one complicating factor, however: Battery lifetimes. Smartphone batteries become unusable after about 2,500 cycles \cite{battery_decline}.
%This is quick enough for it to become significant.
%
Consider a Pixel\,3A with a light-medium usage pattern.
The mean power draw of the device would be 1.54\,W, giving a daily energy consumption of 133\,kJ.
The 3\,Ah (45 \,kJ) battery included in the Pixel would require three full daily charges.
After 833 days or 2.3 years, the battery would be unusable and have to be replaced.

The Pixel 3A's battery has an embodied carbon of 2.00\,kg\ce{CO_2}e and a projected lifespan of 2.3 years.
The Nexus~4's 2.1\,Ah battery and 1.8\,W average power yields an embodied carbon of 1.11\,kg\ce{CO_2}e and a projected lifespan of 1.23 years. 
For any reused smartphone, the embodied carbon cost of replacing the battery $\mathbb{C}_{M}$ is:
\begin{equation}
    \mathbb{C}_{M} = \mathbb{C}_{M(BATTERY)}*\left \lceil{\frac{lifetime}{battery\ lifetime}}\right \rceil 
\end{equation}
This replacement is also costly in terms of human labor. We have replaced a Nexus 4 battery and found it takes about 10 minutes. For a server-equivalent cluster of 54 (Pixel 3As) or 270 (Nexus 4) phones, this would imply 9 (Pixel 3A cluster) - 45 (Nexus 4 cluster) hours of labor every 1.2-2.3 years. We believe this is reasonable; however, this upkeep may become prohibitive for larger systems. Removing and bypassing the battery completely might be worthwhile despite the loss of smart charging and using the battery as a UPS.

\begin{figure*}
    \centering
    \includegraphics[width=\textwidth]{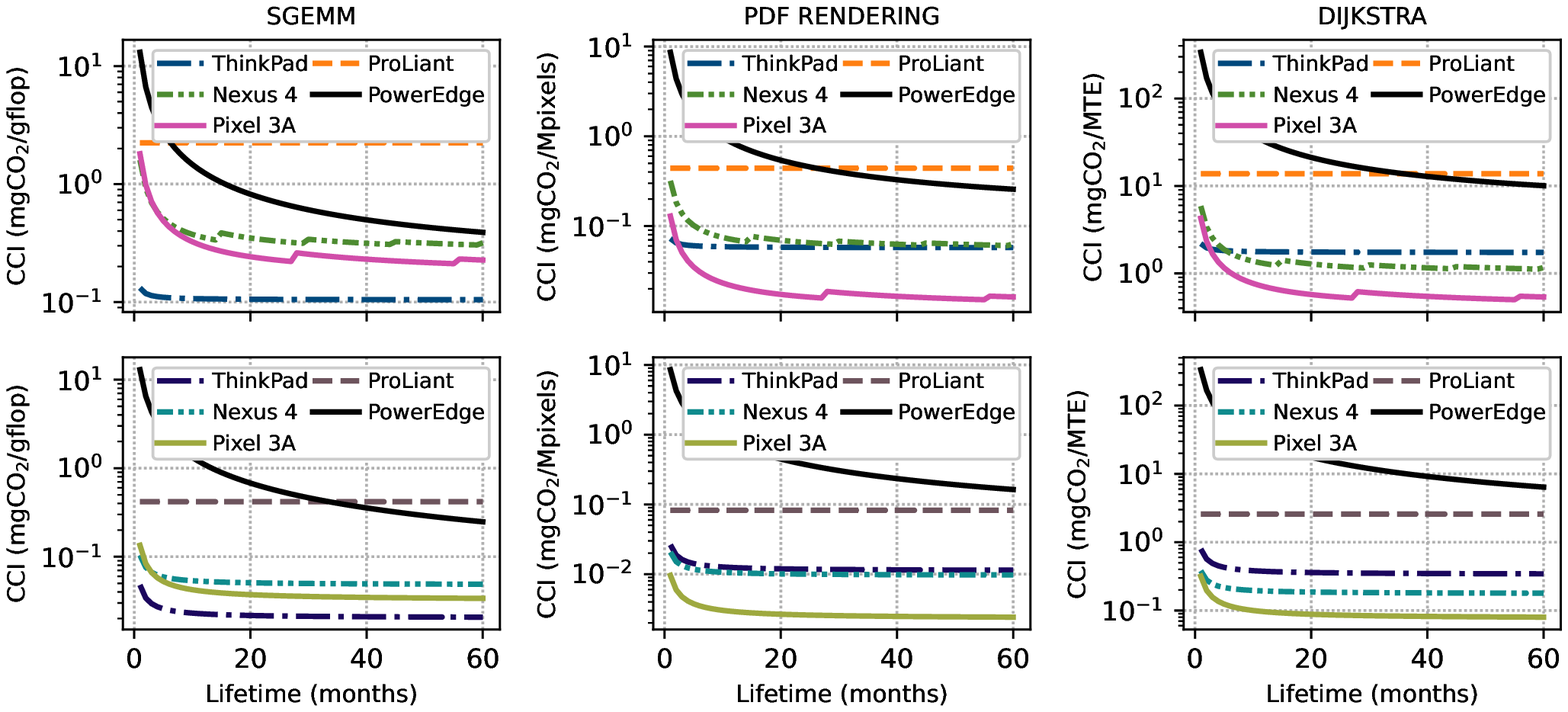}
    \caption{\textbf{Cluster-level CCI for the smartphone clusters vs. an equivalent number of new PowerEdge servers, old Proliant servers, and old ThinkPad laptops}.
    \textmd{The six graphs have three benchmarks with two power regimes. The first row shows CCI for a California energy mix with smart charging for the laptops and smartphones. Smart charging requires a periodic replacement of the battery. The second row assumes access to solar power 100\% of the time, obviating the need for charging and eliminating the batteries.}}
    \label{fig:cci_clusters}
\end{figure*}
\section{Characterizing Carbon}
\label{sec:characterizing-carbon}
\begin{figure}
    \centering
    \includegraphics[width=0.45\textwidth,trim=0.5cm 0.5cm 0.5cm 0.5cm]{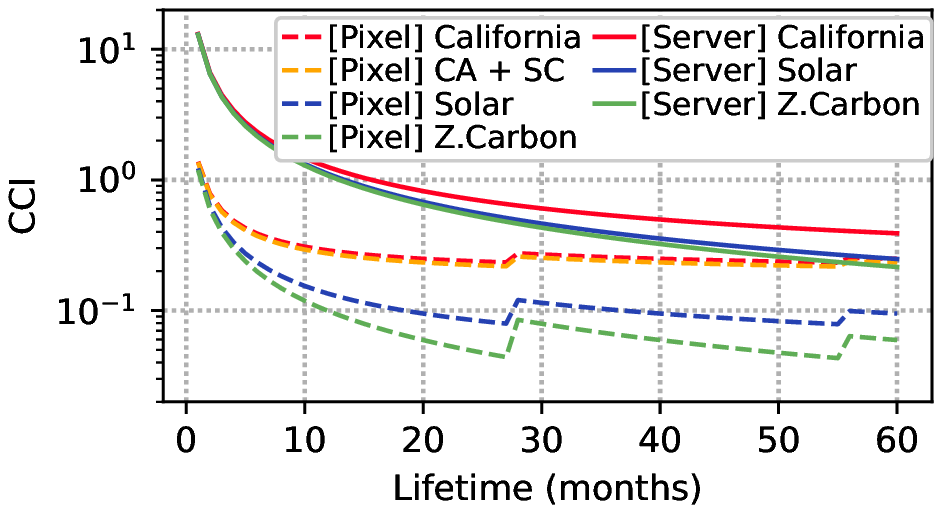}
    \caption{\textbf{The energy mix has a significant impact on CCI.}
    \textmd{The curves show the CCI for the PowerEdge server and the Pixel 3A smartphone running the SGEMM benchmark.}}
    \label{fig:cci_energy}
\end{figure}

We consider the design factors from the previous section to evaluate the carbon savings of junkyard smartphone servers.
\subsection{Sourcing Energy}
First, we consider the energy source's effect on the carbon intensity of new and reused devices. The carbon emissions associated with compute, $\mathbb{C}_{C}$, are equivalent to the carbon footprint of the wallplug energy used to power the device. This can be expressed as:
\begin{equation}
    \mathbb{C}_{C} = \textnormal{CI}_\textnormal{grid} * \sum\limits_{lifetime} P_{avg}
\end{equation}

The carbon intensity of the grid ($\textnormal{CI}_\textnormal{grid}$) is a measure of the amount of \ce{CO_2}e released per kWh of energy provided and varies depending on the source of that energy. For instance, the carbon intensity of solar is 48\,gCO$_{2}$-e/kWh and is 602\,gCO$_{2}$-e/kWh for gas. In California, the mean carbon intensity of grid power is 257\,gCO$_{2}$-e/kWh. \cref{fig:cci_energy} examines how variations in local energy supply affect CCI. Unsurprisingly, more renewable energy means a lower carbon footprint. We can get additional gains with our smart-charging algorithm.

\cref{fig:cci_energy} considers three power regimes. The first is an energy mix that models the California grid. The second is a theoretical regime in which solar energy is always available. While this is not realistic today, hyperscalers have announced plans for 24/7 carbon-free energy within the next decade~\cite{google247green}. 

The third is a theoretical, 100\% carbon-free energy source. Note that such an energy source does not currently exist. Even solar, wind, and hydro have carbon costs associated with their generation, which is reflected in their carbon intensity. However, it provides a theoretical lower bound. In this case, $\mathbb{C}_{C}$ is zero, and as \cref{fig:cci_energy} shows (green lines), $\mathbb{C}_{M}$ becomes the dominant factor for CCI. 

\subsection{Cloudlet-scale Computing}\label{sec:cloudlet}
We now compare the carbon intensities of each previously introduced comparison point at the cloudlet scale. That is, we consider the number of devices and the peripherals necessary to make a cluster of smartphones, laptops, or old servers that meet the performance of a PowerEdge R740 server. We also consider the effect of networking (reflected by the $\mathbb{C}_{N}$ term).
Our cloudlets are:
\begin{enumerate}
    \item A single PowerEdge R740.
    \item 17 Lenovo ThinkPad Gen 3s, with 17 smartplugs added to achieve smart charging savings of 4\%.
    \item 20 ProLiants.
    \item 54 Pixel 3A phones, with 20\% designated as networking and management nodes. 54 smartplugs are added to achieve smart charging savings of 7\%, and one server fan is rated for a 500\,W TDP. 
    \item 256 Nexus 4 phones, with 20\% designated as networking and management nodes. 270 smartplugs are added to achieve smart charging savings of 7\% and two server fans rated for a 500\,W TDP. 
\end{enumerate}
We assume that items \#1-3 are plugged into an already-existent wired network, while items \#4-5 achieve network connectivity via the tree topology described in \Cref{sec:network}

\paragraph{Manufacturing ($\mathbb{C}_{M}$)} The inclusion of hardware peripherals adds a new term to $\mathbb{C}_{M}$: $\mathbb{C}_{M(p)}$, which represents the carbon cost of the added peripherals.  
The cloudlet-scale $\mathbb{C}_{M}$ for the repurposed smartphone- and laptop-based systems is:
\begin{equation}
\smaller
\begin{split}
    \mathbb{C}_{M} = N * \mathbb{C}_{M(batt.)}*\left \lceil{\frac{lifetime}{battery\ lifetime}}\right \rceil  +\ \mathbb{C}_{M(p)}
\end{split}
\end{equation}
where $N$ is the number of devices.
\paragraph{Networking ($\mathbb{C}_{N}$)} We calculate the carbon footprint of networking as described in \cref{eq:net}:
$$\mathbb{C}_{N} = \sum\limits_{lifetime} \textnormal{CI}_\textnormal{grid} * f_{net} * EI_{net}$$

For the smartphones, $EI_{net}$ varies for 3G, 4G, and WiFi. From \cite{network}, we use 5$\,uJ/byte$ for WiFi, and 11$\,uJ/byte$ for LTE. 
The actual value of $f_{net}$ will vary widely depending on the target workload. We assume 0.1\,Gbps for each cloudlet.

\paragraph{Compute ($\mathbb{C}_{C}$)} At the cloudlet scale, we add the operational carbon cost of any added peripherals. For the smartphone-based cloudlets, this will be the added power consumption of the server fans.
\begin{equation}
\smaller
\begin{split}
    \mathbb{C}_{C} = N * \mathbb{C}_{C(device)} + \mathbb{C}_{C(p)}
\end{split}
\end{equation}

The results of this cloudlet-scale calculation are presented in \Cref{fig:cci_clusters}. We plot the results across three benchmarks: SGEMM, PDF Rendering, and Dijkstra, and for two power regimes, California mix and 100\% solar. In the 100\% solar regime, we remove the smartplugs from the equation since they are not needed for smart charging.

In all cases, the repurposed smartphones and laptops perform better than the new server, especially for shorter lifetimes. The old server performs the worst overall due to its relatively high energy consumption ($\nicefrac{2}{3}$ that of the PowerEdge despite being significantly less powerful). For the SGEMM benchmark, the ThinkPad performs the best overall; for the other two benchmarks, the Pixel performs the best. The carbon savings become even more pronounced in the 100\% solar regime, where embodied carbon dominates.

It is interesting to note that the Nexus 4 smartphone cluster, despite consuming more energy (456\,W) than the new PowerEdge (309\,W), is nonetheless still more carbon efficient for both the PDF Rendering and Dijkstra benchmarks. For the SGEMM benchmark, the Nexus 4 cluster is more carbon efficient for lifetimes less than 45 months. In other words, running the higher-powered Nexus 4 phone cluster is better than manufacturing a new server if that server will be in service for less than 45 months. The more efficient Pixel 3A smartphone cluster beats out the server every time.

\subsection{Datacenter-Scale Analysis}

Like PUE, CCI can be calculated at the datacenter scale and provides new insights not reflected by PUE alone.

To illustrate this, we estimate CCI and PUE for a 50MW datacenter built from clusters of Pixel 3A smartphones and compare this against the same built from PowerEdge R740 servers. With the PowerEdge at 308\,W and the Pixel 3A cluster at 84\,W (54 Pixel 3A phones on a light-medium workload), we provision each 50\,MW datacenter with 170,000 units. We assume each smartphone cluster will take up 2U of space, which is enough to leave 75\% of the space empty.

\subsubsection{PUE}
We make the simplifying assumptions of no personnel or windows and define PUE as:
\begin{equation}
PUE = \frac{P_{IT}+ P_{cooling} + P_{lighting}}{P_{IT}}
\end{equation}
We follow the methodology in \cite{pue} to get estimates for $P_{cooling}$ and $P_{lighting}$, using the wattage and size of the PowerEdge and phone cluster. This gives a PUE of 1.32 for the smartphone-based design and 1.31 for the traditional server-based system. The smartphone-based design has a slightly larger PUE because it takes up more physical space, which means that $P_{cooling}$ and $P_{lighting}$ are higher.

\subsubsection{CCI}
For datacenter scale: 
\begin{equation}
    \textnormal{CCI} = \frac{\mathbb{C}_{M}+PUE*(\mathbb{C}_{C}+\mathbb{C}_{N})}{\sum\limits_{lifetime}\textnormal{flop}}
\end{equation}
Using the PUE results from above, and assuming a California energy mix and three-year lifespan, we get the macro-level CCI results given in \Cref{tab:macro_cci}.
\begin{table}[]
    \centering
    \begin{tabular}{c|ccc}
        & \multirow{2}{*}{\textbf{SGEMM}} & \textbf{PDF Render} & \textbf{Dijkstra} \\
        & \footnotesize $\textnormal{mgCO}_{\textnormal{2}}\textnormal{-e}\textnormal{/gflop}$ 
        & \footnotesize $\textnormal{mgCO}_{\textnormal{2}}\textnormal{-e}\textnormal{/Mpixel}$ 
        & \footnotesize $\textnormal{mgCO}_{\textnormal{2}}\textnormal{-e}\textnormal{/MTE}$ \\
        \hline
        \textbf{PowerEdge} & 0.598 & 0.394  & 15.4 \\
        \textbf{Smartphone} & 0.292  & 0.021 & 0.691\\
    \end{tabular}
    \caption{Theoretical three-year datacenter-scale CCI projections for a PowerEdge and smartphone-based system.}
    \label{tab:macro_cci}
\end{table}

\section{Junkyard Cloudlet}
\label{sec:junkyard_cloudlet}
\begin{figure*}
    \centering \includegraphics[width=\textwidth,trim=0cm 0.5cm 0cm 0cm]{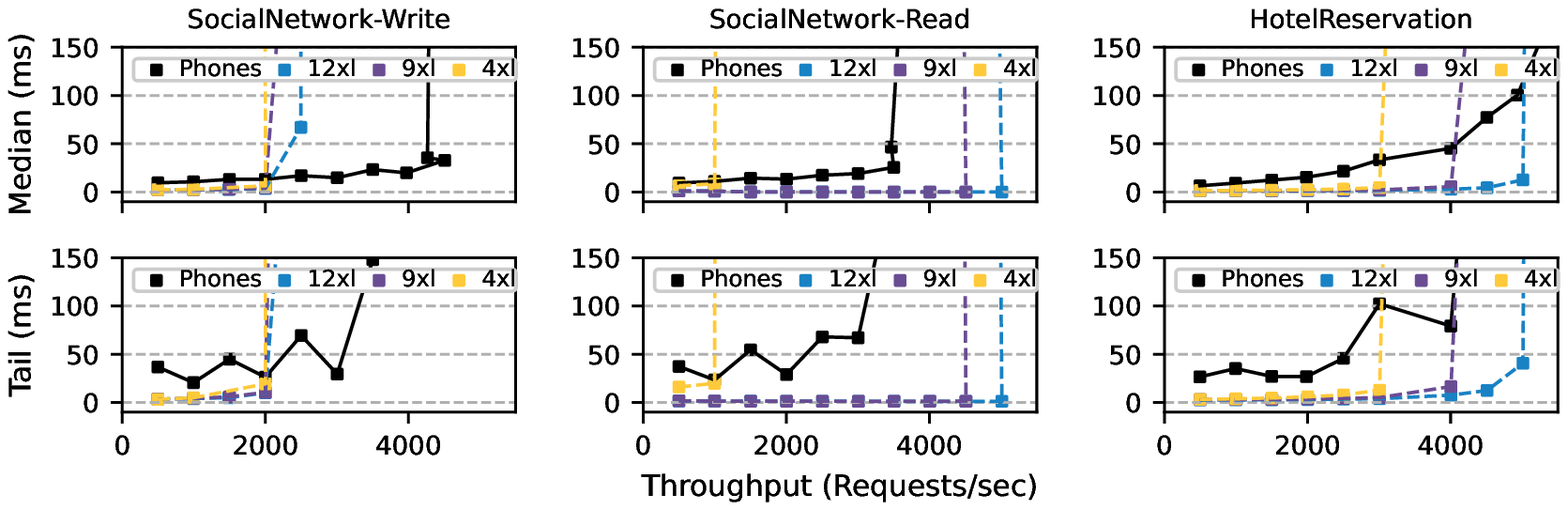}
    \caption{The performance of the smartphone cloudlet vs. different AWS EC2 instances for the three DeathStarBench workloads. \textmd{The top graph shows the median (50\%) latency and bottom graph gives the tail (90\%) latency. For the SocialNetwork-Write, the smartphone cluster outperforms a c5.12xlarge instance; however, for the SocialNetwork-Read, its performance is between c5.4xlarge and c5.9xlarge. For the HotelReservation with its mixed workload, the performance is similar to a c5.9xlarge instance.}}
    \label{fig:deathstar}
\end{figure*}

To demonstrate the feasibility of smartphone-based cloudlet computing we build out a proof-of-concept cluster of ten Pixel 3A smartphones and evaluate our junkyard prototype against a range of similarly-powerful AWS EC2 instances using the DeathStarBench benchmarking suite \cite{deathstarbench}, which provides complete end-to-end microservice-based cloud applications.

\subsection{Experimental Methodology} 
The testbed consists of ten Google Pixel 3A and Pixel 3A XL phones, initially released in 2019. We replace the native Android OS with Ubuntu Touch, an open-source mobile OS that provides a desktop-like experience \cite{ut}. To support Docker, which DeathStarBench relies upon, we further modify the Ubuntu Touch kernel to add several necessary modules, e.g., the BTRFS file system~\cite{btrfs}.

\paragraph{Benchmarks} We implemented two of the three publicly available end-to-end applications from the DeathStarBench suite. They are described below based on information from the DeathStarBench paper and public GitHub~\cite{deathstarbench,dsb_git}. %,nightcore
\begin{itemize}
\item \textit{HotelReservation} is a service that supports getting location-based hotel information and rates, recommending hotels based on user metrics, and placing reservations. It is built with Go and gRPC. The benchmark includes a mixed workload generator. 
% and uses several back-end persistent databases and a recommender system
\item \textit{SocialNetwork} implements unidirectional follow relationships and supports creating text and media posts, reading posts, reading an entire user timeline, searching a database for users or posts, registering, logging in, and following. It is implemented in C++ and uses Thrift RPCs for communication. There are three workload generators: composing posts, reading user timelines, and reading home timelines. We present results for the first two workload generators.
%:  SocialNetwork-Write and SocialNetwork-Read.
\end{itemize}

We attempted to deploy the third public DeathStarBench application (MovieReviewing) on our smartphone cloudlet but found that it did not scale well to multiple devices. The median and tail latency increased with an increasing number of devices. This has been observed by other researchers~\cite{nightcore}, so we do not believe that it is a limitation of the smartphone platform but rather a property of the benchmark itself.

\paragraph{AWS Experiments} As a baseline comparison point, we test each benchmark on differently-sized AWS EC2 C5 instances, described as providing ``cost-effective high performance'' for ``advanced compute-intensive workloads''~\cite{aws_c5}. The applications run on a single AWS machine. The client workload generator operates in another thread on the same machine to eliminate the network latency between our devices and AWS. Since all microservices are deployed on the same node, there is also no network latency between microservice function calls.

\paragraph{Smartphone Testbed} The smartphones are connected over WiFi. They run in Docker Swarm mode, which distributes microservices amongst the phones according to the dependencies defined in the \verb+docker-compose-swarm.yml+ configuration file on the DeathStarBench GitHub~\cite{dsb_git}. To minimize network latency, all phones and the client are on the same local WiFi. Device communication has added latency since the microservices are spread amongst multiple devices.

\subsection{DeathStarBench Performance}
\Cref{fig:deathstar} shows the performance of the phone cloudlet against various AWS EC2 instances. The smartphone cloudlet has a higher tail and median latency across all throughputs; however, it scales well to a large number of requests. This higher latency is expected, given the added network latency between devices in the swarm.

The latency requirements of the system would determine whether the smartphone cloudlet is an acceptable replacement for a cloud server. For instance, if the hotelReservation system required that median latency be no more than 50\,ms and tail latency no more than 100\,ms, the smartphone cloudlet could meet the specs and handle up to 4,000\,queries per second. In this case, the performance of the cloudlet would approximate that of a c5.9xlarge EC2 instance, which costs \$1.53 per hour. 

The smartphone cloudlet costs \$1,027.60\,USD for three years of deployment, which includes the initial \$70\,USD cost per Pixel 3A and California energy prices. The c5.9xlarge EC2 instance costs \$40,404\,USD (\$1.53\,USD per hour) for the same deployment length.

Some interesting differences exist between the smartphone cluster performance on the different benchmarks. Normalized to the AWS equivalents, the smartphone cluster performs significantly better on the write-only SocialNetwork application than on the read-only application. This may be explained by the fact that the read workload involves the transfer of a user's entire timeline – a large amount of data.
\begin{figure}
    \centering
    \includegraphics[width=0.5\textwidth,trim=0.5cm 0.5cm 0.5cm 0.5cm]{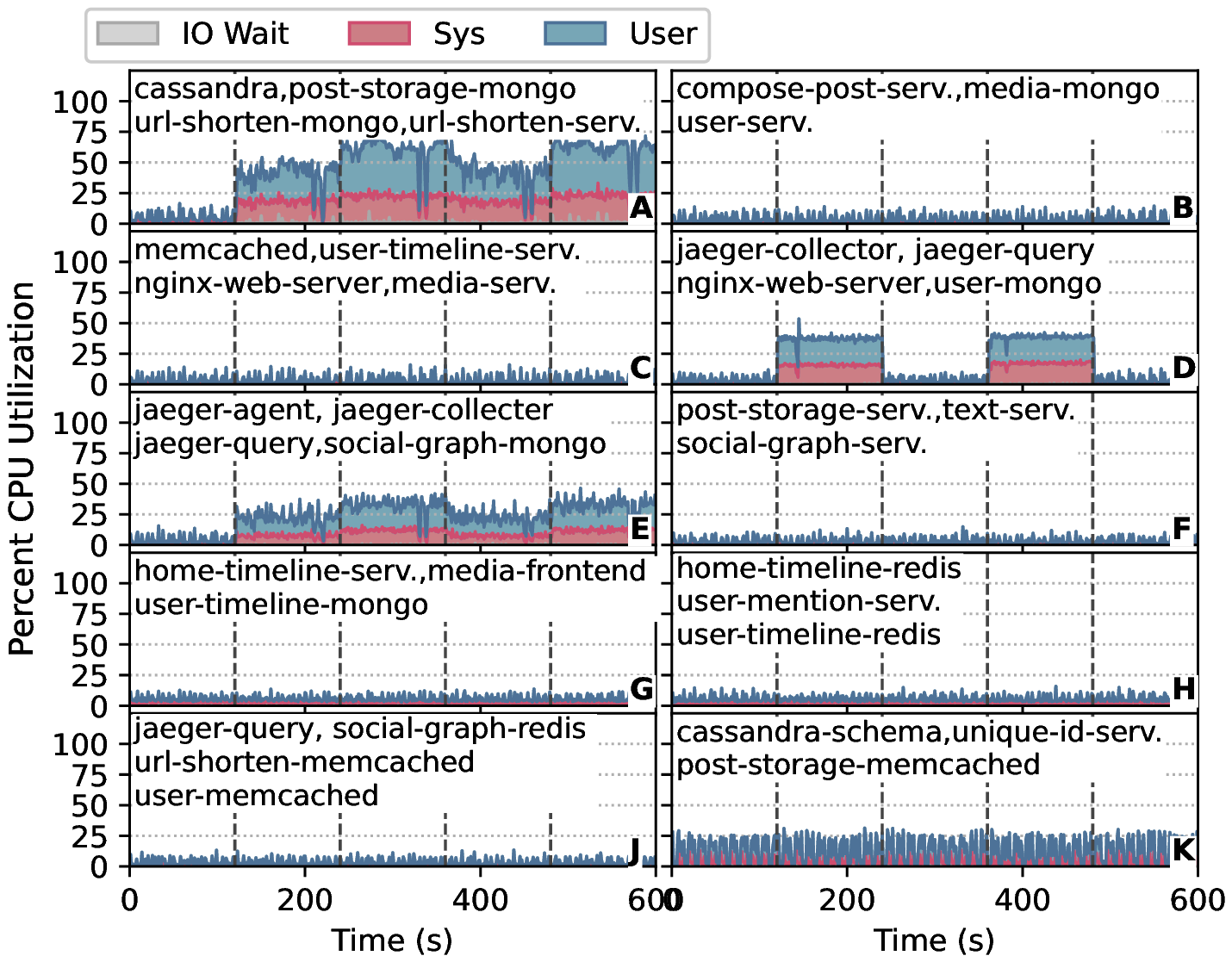}
    \caption{CPU utilization for each of the ten smartphones running the SocialNetwork benchmark. \textmd{The system is idle from 0-120s. The SocialNetwork-Read workload is run at 3,000 QPS from 120-240s. No requests are submitted from 240-360s. The SocialNetwork-Write workload is run at 3,500 QPS from 360-480s. The system is idle again from 480-600s. The vertical dotted lines indicate the starting/stopping points. The microservices hosted by each device are shown on the graph.}}
    \label{fig:cpus}
\end{figure}

\paragraph{CPU Utilization} The CPU utilization of the smartphones (\Cref{fig:cpus}) indicates two interesting points. First, the smartphones were generally not CPU-bound while hosting these microservices. Second, the utilization varied widely depending on the services being hosted by the device, with 6/10 of the devices making little use of their raw compute powers.
%(B, C, F, G, H, J). 

We ran the same experiment on the c5.9xlarge and found that it was similarly not CPU-bound: the CPU utilization was relatively constant at 30\% utilization while running the SocialNetwork-Read workload at 4,500 QPS and 25\% while running the SocialNetwork-Write workload at 2,000 QPS. 

\subsection{Carbon Performance}

\begin{figure}
    \centering
    \includegraphics[width=0.5\textwidth]{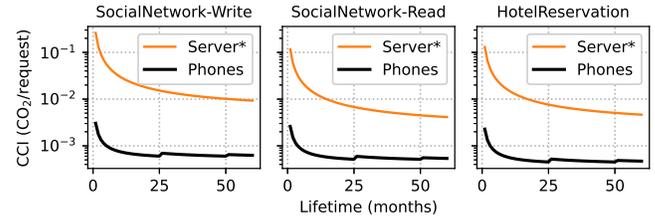}
    \caption{Carbon intensity of Pixel 3A cloudlet running the DeathStarBench applications compared to c5.9xlarge server.}
    \label{fig:cci_ult}
\end{figure}

A Pixel 3A consumed approximately 1.7\,W while hosting the hotel reservation system. Therefore, we assume that the ten-device system running at 17\,W requires, at most, a single server fan. With this energy consumption, the phones would have to be charged 3.4 times a day, and batteries would have to be replaced every 2.1 years. Using the CCI methodology, \Cref{fig:cci_ult} shows the carbon intensity of the Pixel 3A cloudlet. 

It is impossible to know the actual power consumption of the EC2 instances without having access to their custom hardware. We use the public dataset of rough estimates based on benchmarking very similar hardware \cite{aws_dataset} to approximate that a c5.9xlarge instance will use 140.7\,W at 10\% utilization, and 239\,W at 50\%. The CPU benchmarking indicates that the c5.9xlarge will likely be at 25-30\% utilization while hosting SocialNetwork; for simplicity, we round down to 10\% and use the 140.7\,W as an estimate for c5.9xlarge. The same dataset estimates the manufacturing carbon of the c5.9xlarge as 1344\,kg\,CO$_2$-equivalent.

The throughput of each alternative is determined by considering the point on the curves in \Cref{fig:deathstar} at which throughput is at its max before the latencies shoot up. For the smartphone cluster, this is 4,000 QPS for HotelReservation, 3,000 QPS for SocialNetwork-Write, and 3,500 QPS for SocialNetwork-Read. The throughput for C5.9xlarge is 4,000 QPS for HotelReservation, 4,500 QPS for SocialNetwork-Read, and 2,000 QPS for SocialNetwork-Write.

\Cref{fig:cci_ult} shows the final result. For each benchmark studied, the smartphone-based system is significantly more carbon efficient, in terms of CO$_{2}$ per query. After three years of use, the relative carbon efficiency of the smartphone cluster vs. the c5.9xlarge singlet is 18.9x more carbon efficient t for SocialNetwork-Write, 9.8x more carbon efficient for SocialNetwork-Read, and 12.6x more carbon efficient for HotelReservation with its mixed workload.

\section{Conclusion}
\label{sec:conclusion}
The major takeaways of this work are:
\begin{enumerate}
    \item For specific workloads, clusters of repurposed phones are cheaper and more carbon efficient than traditional servers.
    \item More broadly, scavenging unwanted equipment shows excellent potential for building economic and carbon-efficient systems, especially when renewable energy is plentiful.
    \item Sustainability has operational and manufacturing facets; manufacturing dominates as operating trends towards zero with cleaner energy mixes.
    \item Accurate LCA information is essential for carbon-based analyses; it would be beneficial if more ICT manufacturers published this information, including cloud providers who build custom systems.
\end{enumerate}

\noindent Our work highlights the need for more holistic analyses of the environmental impact of computing.  With the substantial carbon cost of manufacturing and the difficulties of responsible recycling, the energy efficiency of a device may be the least significant component of its environmental and human impact.

%%%%%%%%% -- BIB STYLE AND FILE -- %%%%%%%%
\bibliographystyle{ACM-Reference-Format}
\bibliography{refs}
%%%%%%%%%%%%%%%%%%%%%%%%%%%%%%%%%%%%

\end{document}